\documentclass[11pt,english,longbibliography]{article}

\usepackage{amsthm}

\makeatletter
\newcommand*{\rom}[1]{\expandafter\@slowromancap\romannumeral #1@}
\makeatother

\usepackage[latin9]{inputenc}
\usepackage{geometry}
\usepackage{verbatim}
\usepackage{prettyref}
\usepackage[numbers,sort&compress]{natbib}
\usepackage{amsmath}
\usepackage{amssymb}
\usepackage{tikz-cd}
\tikzset{commutative diagrams/row sep/huge=4cm}
\tikzset{commutative diagrams/column sep/huge=4cm}
\tikzcdset{scale cd/.style={every label/.append style={scale=#1},
    cells={nodes={scale=#1}}}}
    \tikzset{
    Witten diagram/.style={
        execute at begin picture={
            \draw[blue, line width=1.5pt] circle[radius=\pgfkeysvalueof{/tikz/Witten/radius}];
            \path node (X){\phantom{X}};
        },
        baseline={(X.base)}
    },
    vertex/.style={circle,fill,inner sep=1.5pt,node contents={}},
    Witten/.cd,
    radius/.initial=3cm
}
\usepackage{lmodern}
\usepackage{tcolorbox}
\usepackage{cases}
\usepackage{bbold}
\usepackage{bm}
\usepackage{ytableau}
\usepackage{youngtab}
\usepackage{mathtools}
\usepackage{graphicx}
\usepackage[nottoc]{tocbibind}
\usepackage{mathrsfs}
\usepackage{caption}
\usepackage{subcaption}
\usepackage{cancel}
\usepackage{float}
\usepackage{tikz}
\usepackage{lipsum}
\usepackage{framed}
\usepackage{adjustbox}
\usepackage{mathtools}
\usepackage{braket}
\usepackage[compat=1.1.0]{tikz-feynman}
\usetikzlibrary{decorations.markings}
\usetikzlibrary{arrows.meta}
\usetikzlibrary{shapes.geometric}
\usetikzlibrary{calc}

\usepackage{slashed}
\usepackage{esvect}
\usepackage{dsfont}
\usepackage{color}
\definecolor{darkgreen}{rgb}{0,0.5,0}
\definecolor{darkblue}{rgb}{0,0,0.6}
\definecolor{purple}{rgb}{0.4,.2,0.7}
\usepackage[colorlinks=true,citecolor=darkgreen,linkcolor=purple,urlcolor=purple]{hyperref}

\numberwithin{equation}{section}
\numberwithin{figure}{section}
\numberwithin{table}{section}

\DeclareFontShape{OT1}{cmr}{mx}{n}{<->cmr10}{}

\begin{document}

\fontseries{mx}\selectfont

\begin{center}
{\LARGE \bf 
Towards a Quintic Ginzburg-Landau Description \\
of the $(2,7)$ Minimal Model 
\\[6ex]}
{\Large Andrei Katsevich,\textsuperscript{$\phi$} Igor Klebanov,\textsuperscript{$\phi$, $\phi^2$} Zimo Sun,\textsuperscript{$\phi$, $\phi^3$}
Grigory Tarnopolsky\textsuperscript{$\phi^5$}
\\[2ex]}
    {\normalsize \textsuperscript{$\phi$}Joseph Henry Laboratories, Princeton University, Princeton, NJ 08544\\}
     {\normalsize \textsuperscript{$\phi^2$}Princeton Center for Theoretical Science, Princeton University, Princeton, NJ 08544\\}
      {\normalsize \textsuperscript{$\phi^3$} Institute for Advanced Study, Princeton, NJ 08540\\}
{\normalsize \textsuperscript{$\phi^5$} Department of Physics, Carnegie Mellon University, Pittsburgh, PA 15213 \\}
 
\end{center}
\vskip2cm

\begin{abstract}
We discuss dimensional continuation of the massless scalar field theory with the $i\phi^5$ interaction term. It preserves the so-called $\mathcal{PT}$ symmetry, which acts by $\phi\rightarrow -\phi$ accompanied by $i\rightarrow -i$. Below its upper critical dimension $10/3$, this theory has interacting infrared fixed points. We argue that the fixed point in $d=2$ describes the non-unitary minimal conformal model $M(2,7)$. We identify the operators $\phi$ and $\phi^2$ with the Virasoro primaries $\phi_{1,2}$ and $\phi_{1,3}$, respectively, and $i\phi^3$ with a quasi-primary operator, which is a Virasoro descendant of $\phi_{1,3}$. Our identifications appear to be consistent with the operator product expansions and with considerations based on integrability. Using constrained Pad\' e extrapolations, we provide estimates of the critical exponents in $d=3$. We also comment on possible lattice descriptions of $M(2,7)$ and discuss RG flows to and from this CFT. Finally, we conjecture that the minimal models $M(2, 2n+1)$ are described by the massless scalar field theories with the $i\phi^{2n-1}$ interaction terms.
\end{abstract}

\newpage

\section{Introduction}
Massless scalar field theories can describe critical phenomena in various dimensions. A textbook example is the Euclidean $\phi^4$ field theory, which describes the critical Ising model in dimensions $1<d<4$. In $d=2$, this strongly interacting field theory becomes equivalent to the minimal conformal model $M(3,4)$ \cite{Belavin:1984vu}. The $d$-dependence of the scaling dimensions and OPE coefficients can be studied using various methods, such as the $4-\epsilon$ expansion \cite{Wilson:1971dc} and the conformal bootstrap \cite{Rattazzi:2008pe}, and the results are in excellent agreement with each other \cite{Cappelli:2018vir,Henriksson:2022gpa}. More generally, the massless Euclidean field theories with $\phi^{2n}$ interactions are equivalent in $d=2$ with the $A$-series unitary minimal models $M(n+1, n+2)$ \cite{Zamolodchikov:1986db}. In particular, for the $\phi^6$ field theory, which describes the tri-critical Ising model $M(4,5)$, the $d$-dependence of observables was studied in \cite{Henriksson:2025kws} using a combination of $3-\epsilon$ expansions and conformal bootstrap.

Similar correspondences between Ginzburg-Landau (GL) scalar field theories and critical phenomena also apply in some non-unitary cases. The original such example is the massless $i\phi^3$ Euclidean field theory, which describes the Yang-Lee universality class \cite{Fisher:1978pf}. This field theory possesses the so-called $\mathcal{PT}$ symmetry, which acts by $i\rightarrow - i, \phi \rightarrow - \phi$. The $6-\epsilon$ expansion of the $i\phi^3$ theory was introduced in \cite{Fisher:1978pf} and in recent years has been carried out to much higher orders \cite{Kompaniets:2021hwg,Borinsky:2021jdb,Schnetz:2025wtu}. In $d=2$, this theory becomes equivalent to the minimal model $M(2,5)$ \cite{Cardy:1985yy} , and $\phi$ is identified with the only non-trivial Virasoro primary operator $\phi_{1,2}$.\footnote{A cubic GL description with two scalar fields applies to the $M(3,8)$ and $M(3,10)$ minimal models \cite{Klebanov:2022syt,Katsevich:2024jgq}.} The operator $\phi^2$ is a conformal descendant, while the higher powers of $\phi$ should be identified with operators in $M(2,5)$ that are {\it quasi-primary}, i.e. primary only under the conformal group $SL(2,\mathbb{R})\times SL(2,\mathbb{R})$. In particular, $i\phi^3$ has been identified \cite{ArguelloCruz:2025zuq} with the lowest dimension quasi-primary operator $T\bar T$. Imposing the $d=2$ constraints improves the Pad\'e extrapolations of the $6-\epsilon$ expansions \cite{ArguelloCruz:2025zuq,Gracey:2025rnz}. The resulting estimates are in good agreement with the high-temperature expansions \cite{Butera:2012tq} and with numerical calculations of quantum criticality of the Ising model in an imaginary magnetic field \cite{ArguelloCruz:2025zuq,Fan:2025bhc,EliasMiro:2025msj}, which were pioneered in $d=2$ by von Gehlen \cite{vonGehlen:1991zlm}.

In this paper, we will be primarily concerned with the analogous $\mathcal{PT}$-symmetric quintic theory
\begin{equation}\label{eq:GL27}
    S=\int d^dx\left(\frac{1}{2}(\partial_\mu\phi)^2+\frac{g\phi^5}{120}\right),\quad g\in i\mathbb{R}\,
\end{equation}
and will argue that in $d=2$ it should be identified with the $M(2,7)$ minimal model. An early proposal for the Ginzburg-Landau description of $M(2,7)$ \cite{Becker:1991nr,Becker:1991nq} also involved an $i\phi^5$ interaction term, but the derivative term was unconventional.\footnote{After defining $\chi=\phi^2$, the action density becomes $\frac{1}{2}(\partial_\mu\chi)^2+ig \chi^{\frac{5}{2}}$.} Since then, there have been a number of works on this subject including \cite{vonGehlen:1994rp,Mossa:2007fx,Zambelli:2016cbw,Codello:2017epp,amslaurea11308,Lencses:2022ira,Lencses:2023evr,Lencses:2024wib,Katsevich:2024jgq}, but some disagreements between them have persisted, and in some of them the theory (\ref{eq:GL27}) was identified with $M(2,9)$ instead of $M(2,7)$. The lattice models proposed for $M(2,7)$ have involved the Blume-Capel model \cite{Blume:1966zz,CAPEL1966966} with an added imaginary magnetic field \cite{vonGehlen:1994rp,PhysRevLett.84.4794}, and we will make some comments about them.

The main feature of our new argument in favor of the identification of (\ref{eq:GL27}) with $M(2,7)$ is that only the operators $\phi$ and $\phi^2$ become Virasoro primaries in $d=2$. We identify them with $\phi_{1,2}$ and $\phi_{1,3}$, respectively, and their OPEs agree with the semiclassical treatment of the quintic theory (\ref{eq:GL27}). Importantly, we identify the operator $i\phi^3$ with the {\it quasi-primary} operator $\left(L_{-2}-\frac{21}{2}L_{-1}^2\right) \left(\bar L_{-2}-\frac{21}{2} \bar L_{-1}^2\right) \phi_{1,3}$.

Using the $\frac{10}{3}-\epsilon$ expansions developed for (\ref{eq:GL27}) in \cite{Gracey:2017okb,Codello:2017epp} and imposing the $M(2,7)$ boundary conditions in $d=2$, we will provide Pad\'e estimates of the $d$-dependence of operator dimensions in this universality class. In $d=3$, our estimates of the critical exponents can be hopefully compared with numerical results in appropriate lattice approaches to $M(2,7)$. We also discuss RG flows $M(4,5)\rightarrow M(2,7)$ and $M(2,7)\rightarrow M(2,5)$, which provide further checks of our arguments.

\section{Matching the GL description with $M(2,7)$}
Let us describe the $M(2,7)$ minimal model in detail. It has central charge $c(2,7)=-\frac{68}{7}$ and effective central charge \cite{Itzykson:1986pk,Castro-Alvaredo:2017udm} $c_{\text{eff}}(2,7)=\frac{4}{7}$. This model contains two non-trivial primary operators $\phi_{1,2}$ and $\phi_{1,3}$. Our proposal for operator identifications is summarized in Table. \ref{tab:2}.
\begin{table}[h!]
    \centering
    \begin{tabular}{|c|c|c|c|c|c|c|}
        \hline
        $M(2,7)$ & $\phi_{1,1}$ & $\phi_{1,2}$ & $\phi_{1,3}$ & $ Q_2\bar{Q}_2\phi_{1,3}$ & $ T\bar{T}$ & $Q_3 \bar Q_3 \phi_{1,2}$ \\\hline
        $\Delta $ & $0$ & $-\frac{4}{7}$ & $-\frac{6}{7}$ & $\frac{22}{7}$ & $4$ & $\frac{38}{7}$ \\\hline
        $\mathcal{PT}$ & even & odd & even & even & even & odd \\\hline
        GL & 1 & $\phi$ & $\phi^2$ & $i\phi^3$ & $i\phi^5$ & $i\phi^6$ \\\hline
    \end{tabular}
    \caption{Properties of the primary and some quasi-primary scalar operators of the $M(2,7)$ minimal model.}
    \label{tab:2}
\end{table}

The $\mathcal{PT}$-odd operator $\phi_{1,2}\sim\phi$ has negative scaling dimension $\Delta_{1,2}=2h_{1,2}=-\frac{4}{7}$. The $\mathcal{PT}$-even operator $\phi_{1,3}\sim\phi^2$ has dimension $\Delta_{1,3}=-\frac{6}{7}$. We identify the scalar operators $i\phi^3$, $i\phi^5$, and $i\phi^6$ with quasi-primary Virasoro descendants\footnote{The operator $i\phi^4$ is a conformal descendant by the equation of motion, and in 2D it becomes $L_{-1}\bar{L}_{-1}\phi_{1,2}\equiv\Box\phi$.} 
\begin{equation}
   i\phi^3\sim Q_2\bar{Q}_2\phi_{1,3}\,, \qquad i\phi^5\sim T\bar{T}\,, \qquad i\phi^6 \sim Q_3 \bar Q_3 \phi_{1,2} \,,
\end{equation}
where $Q_2 = L_{-2}-\frac{21}{2}L_{-1}^2$ and $Q_3=L_{-3}+ 7L_{-2}L_{-1}-\frac{49}{10}L_{-1}^3$. Thus, in 2D the operator $i\phi^3$ is irrelevant, while it is relevant in $d=\frac{10}{3}-\epsilon$. We note, however, that in 2D there is a relevant spin-$2$ quasi-primary operator of dimension $\frac{8}{7}$,
\begin{equation}\label{eq:spin2}
   T'_{zz}= Q_2 \phi_{1,3}  \ , \qquad  T'_{\bar z\bar z}=\bar{Q}_2 \phi_{1,3}\ .
\end{equation}
In the GL description, the corresponding operator is $T'_{\mu \nu}= \phi T_{\mu \nu}$.

The OPEs of Virasoro primary operators are given by
\begin{equation}\label{eq:OPE}
    \begin{aligned}
        &\phi_{1,2}\times\phi_{1,2}=1+\phi_{1,3}\,,\\
        &\phi_{1,2}\times\phi_{1,3}=\phi_{1,2}+i\phi_{1,3}\,,\\
        &\phi_{1,3}\times\phi_{1,3}=1+i\phi_{1,2}+\phi_{1,3}\,.
    \end{aligned}
\end{equation}
The non-vanishing three-point functions of primary operators, which are implied by the OPE, are $\langle \phi_{1,2}\,\phi_{1,2}\,\phi_{1,3}\rangle$, $\langle \phi_{1,3}\,\phi_{1,3}\,\phi_{1,2}\rangle$, and $\langle \phi_{1,3}\,\phi_{1,3}\,\phi_{1,3}\rangle$. Using the dictionary in Table \ref{tab:2}, we can reproduce them using the following Feynman graphs.
\begin{equation}
\begin{aligned}
    &\langle \phi_{1,2}\,\phi_{1,2}\,\phi_{1,3}\rangle =
    \begin{tikzpicture}[baseline={(0,-0.1)},]
        \node at (-1.2,0.6) {$\phi$};
        \node at (-1.2,-0.6) {$\phi$};
        \node at (0.3,0) {$\phi^2$};
        \draw (-1,0.5) -- (0,0) -- (-1,-0.5);
    \end{tikzpicture},\quad 
    \langle \phi_{1,3}\,\phi_{1,3}\,\phi_{1,2}\rangle =
    \begin{tikzpicture}[baseline={(0,-0.1)}]
        \node at (-0.2,0.6) {$\phi^2$};
        \node at (-0.2,-0.6) {$\phi^2$};
        \node at (2,0) {$\phi$};
        \draw (0,0.5)  to[out=0,in=130] (1,0) to[out=180,in=-60] (0,0.5);
        \draw (0,-0.5) to[out=0,in=-130] (1,0) to[out=180,in=60] (0,-0.5);
        \draw (1,0) node[vertex]{} -- (1.8,0) ;
        \node at (1,-0.3) {$i$};
    \end{tikzpicture}\,,\\
    &\langle \phi_{1,3}\,\phi_{1,3}\,\phi_{1,3}\rangle =
    \begin{tikzpicture}[baseline={(0,-0.1)}]
        \node at (-1.3,0.6) {$\phi^2$};
        \node at (-1.3,-0.6) {$\phi^2$};
        \node at (0.3,0) {$\phi^2$};
        \draw (-1,0.5) -- (0,0) -- (-1,-0.5) -- (-1,0.5);
    \end{tikzpicture}\,.
\end{aligned}
\end{equation}
The three-point functions of the lowest quasi-primary operator, $\phi^3$, with two Virasoro primaries are given by the diagrams 
\begin{equation}
\begin{aligned}
    &\langle \phi_{1,2}\,\phi_{1,2}\, \phi^{3}\rangle=
    \begin{tikzpicture}[baseline={(0,-0.1)}]
        \node at (-1.2,0.6) {$\phi$};
        \node at (-1.2,-0.6) {$\phi$};
        \node at (1.3,0) {$\phi^3$};
        \node at (0,-0.3) {$i$};
        \draw (-1,0.5) -- (0,0) node[vertex]{} -- (-1,-0.5);
        \draw (1,0)--(0,0);
        \draw (1,0) to[out= 150,in= 30] (0,0);
        \draw (1,0) to[out=-150,in=-30] (0,0);
    \end{tikzpicture},\quad\langle \phi_{1,2}\,\phi_{1,3}\, \phi^{3}\rangle=
    \begin{tikzpicture}[baseline={(0,-0.1)}]
        \node at (-1.2,0.6) {$\phi$};
        \node at (-1.2,-0.6) {$\phi^2$};
        \node at (0.4,0) {$\phi^3$};
        \draw (-1,0.5) -- (0,0);
        \draw (-1,-0.5) to[out=0,in=-130] (0,0) to[out=180,in=60] (-1,-0.5);
    \end{tikzpicture}\,,\\
    &\langle\phi_{1,3}\,\phi_{1,3}\, \phi^{3}\rangle=
    \begin{tikzpicture}[baseline={(0,-0.1)}]
        \node at (-1.2,0.6) {$\phi^2$};
        \node at (-1.2,-0.6) {$\phi^2$};
        \node at (1.3,0) {$\phi^3$};
        \node at (0,-0.3) {$i$};
        \draw (-1,0.5) -- (0,0) node[vertex]{} -- (-1,-0.5) -- (-1,0.5);
        \draw (1,0)--(0,0);
        \draw (1,0) to[out= 150,in= 30] (0,0);
        \draw (1,0) to[out=-150,in=-30] (0,0);
    \end{tikzpicture}\,.\\
\end{aligned}
\end{equation}
These perturbative explanations of the OPE are a good test of our GL interpretation of $M(2,7)$. They can be considered as generalizations of Zamolodchikov's OPE-based argument for the GL description of unitary minimal models $M(n,n+1)$ \cite{Zamolodchikov:1986db}.

\begin{figure}[t]
    \centering
    \includegraphics{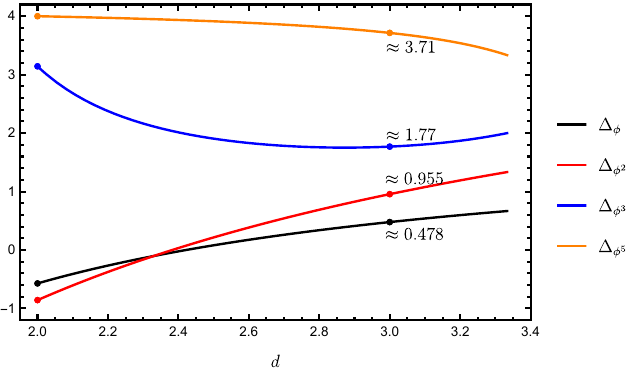}
    \caption{Two-sided Pad\'e extrapolations for the dimensions of operators $\phi, \phi^2, \phi^3$ and $\phi^5$. We use the [1,1] Pad\'e approximant, except for $\phi^3$ where we use the [0,2] Pad\'e approximant because [1,1] has a pole. }
    \label{Padeplot}
\end{figure}

The quintic theory (\ref{eq:GL27}) is weakly coupled in $d=\frac{10}{3}-\epsilon$  dimensions. Its beta function was computed up to the $g^3$ order \cite{Gracey:2017okb,Codello:2017epp}. Rescaling $g$ as in \cite{Codello:2017epp}, the beta function reads
\begin{equation}
    \beta=-\frac{3}{2}g\epsilon-\frac{1377}{16}g^3+\mathcal{O}(g^5)\,,
\end{equation}
which admits a pair of imaginary fixed points $g_*\propto \pm i\sqrt{\epsilon}$. Using the anomalous dimensions calculated in \cite{Gracey:2017okb,Codello:2017epp}, we obtain the scaling dimensions of $\phi, \phi^2,\phi^3$ and $\phi^5$ at the imaginary fixed points to order $\epsilon$:
\begin{equation}\label{epsexp}
    \begin{aligned}
        &\Delta_\phi=\frac{2}{3}-\frac{383\epsilon}{765}+\mathcal{O}(\epsilon^2),\quad\Delta_{\phi^2}=\frac{4}{3}-\frac{262\epsilon}{255}+\mathcal{O}(\epsilon^2)\,,\\
        &\Delta_{\phi^3}=2-\frac{313\epsilon}{255}+\mathcal{O}(\epsilon^2),\quad\Delta_{\phi^5}=\frac{10}{3}+2\epsilon+\mathcal{O}(\epsilon^2)\,.
    \end{aligned}
\end{equation}
Based on the operator identification in Table \ref{tab:2}, we know the exact scaling dimensions of these operators in $d=2$ dimensions. This information allows us to apply two-sided Pad\'e resummations to the $\epsilon$-expansions in \eqref{epsexp}. In Fig. \ref{Padeplot}, we show the resulting Pad\'e extrapolations for $2\le d\le \frac{10}{3}$, including the rough estimates of the scaling dimensions in $d=3$. The corresponding rough estimates of the critical exponents are $\eta=2\gamma_\phi \approx -0.044$, $\nu =\frac{1}{d-\Delta_{\phi^2}}\approx 0.49$, $\zeta=\frac{d-\Delta_{\phi^3}}{d-\Delta_\phi} \approx 0.49$, $\sigma = \frac{\Delta_\phi}{d-\Delta_\phi}\approx 0.19$, $\omega = \Delta_{\phi^5}-d \approx 0.71$. For comparison, the exact critical exponents in $d=2$ are
$\eta=-\frac{8}{7}$, $\nu=\frac{7}{20}$, $\zeta=-\frac{4}{9}$, $\sigma=-\frac{2}{9}$, $\omega=2$. 

As $d$ is reduced from $10/3$, the scaling dimensions of $\mathcal{PT}$-odd operator $\phi$ and $\mathcal{PT}$-even operator $\phi^2$ cross at $2<d_*<3$. Since these operators are in different $\mathcal{PT}$ sectors, there is no symmetry-based obstruction to a genuine crossing. We defer a quantitative study (e.g. locating $d_*$ and testing for exceptional-point behavior \cite{Heiss:2012dx}) to future work.

\section{Integrable deformation by $\phi_{1,3}$}\label{sec:3}
Deforming the $M(2,7)$ model by $\phi_{1,3}\sim \phi^2$ yields a massive integrable theory, whose spectrum  consists of two breathers $A_a$, $a\in\{1,2\}$ with the golden ratio between masses \cite{Smirnov:1990vm,Freund:1989jq}:
\begin{equation}
    \frac{m_2}{m_1}=2\cos\frac{\pi}{5}\approx1.68103\,.
\end{equation}
There are no kinks in this theory. The two-particle $S$-matrix of the corresponding reduced sine-Gordon model with $\beta^2=\frac{16\pi}{7}$ is \cite{Smirnov:1990vm,Freund:1989jq}
\begin{equation}
    \begin{aligned}
        S_{11}(\theta)=F_{\frac{2}{5}}(\theta),\quad S_{12}(\theta)=F_{\frac{3}{5}}(\theta)F_{\frac{4}{5}}(\theta),\quad S_{22}(\theta)=F^2_{\frac{3}{5}}(\theta)F_{\frac{4}{5}}(\theta)\,,
    \end{aligned}
\end{equation}
where $\theta$ is the rapidity difference of the particles and the building block $F_\alpha(\theta)$ is a meromorphic function of $\theta$
\begin{equation}
    F_\alpha(\theta)=\frac{\tanh(\frac{1}{2}(\theta+i\pi\alpha))}{\tanh(\frac{1}{2}(\theta-i\pi\alpha))}\,.
\end{equation}
When $0<\alpha<\frac{1}{2}$, $F_\alpha(\theta)$ has a simple pole of residue $2i\tan\alpha\in i\mathbb{R}_+$ at $i\alpha\pi$ and a simple pole of residue $-2i\tan\alpha\in i\mathbb{R}_-$ at $i(1-\alpha)\pi$. $F_{\frac{1}{2}}(\theta)$ has only a double pole at $\frac{i\pi}{2}$. The  $S$-matrix satisfies the spatial reflection symmetry $S_{ab}(\theta)=S_{ba}(\theta)$, the crossing symmetry and unitarity respectively \cite{Zamolodchikov:1989hfa}.

In the deformed  theory $M(2,7)+\phi_{1,3}$, the breather $A_1$ exhibits a ``$\phi^5$-property", with the scattering processes given by $A_1A_1\rightarrow A_2$, $A_1A_2\rightarrow A_1$, $A_1A_2\rightarrow A_2$, $A_2A_2\rightarrow A_1$\footnote{Analogously, the Scaling Lee-Yang Model (SLYM) $M(2,5)+i\phi_{1,3}$ has a fundamental particle $A_1$ satisfying the ``$\phi^3$-property": $A_1A_1\rightarrow A_1$. This amplitude is purely imaginary \cite{Cardy:1989fw} (see also \cite{Zamolodchikov:1989hfa} and \cite{Cardy:2023lha}).}. These scattering processes can be represented diagrammatically as follows:
\begin{equation}\label{fabc}
\vcenter{\hbox{\begin{tikzpicture}[scale=1, baseline={(0,-0.1)}, rotate=-90]
    \draw[thick, blue] (0,0) -- (0,1.2);
    \draw[thick, red] (0,0) -- ++(234:1.2); 
    \draw[thick, red] (0,0) -- ++(-54:1.2); 
    \draw [thick, <->, >=stealth] (0.0436,0.498) arc[start angle=85,end angle=-50, radius=0.5];
    \draw  [thick, <->,>=stealth]  (-0.0436, 0.498) arc[start angle=95,end angle=230, radius=0.5];
    \draw  [thick, <->, >=stealth]  (0.265,-0.424) arc[start angle=-58,end angle=-122, radius=0.5];
    \node[transform shape=false] at (0.7,0.2) {\tiny $\frac{4\pi}{5}$};   
    \node[transform shape=false] at (-0.7,0.2) {\tiny $\frac{4\pi}{5}$};  
    \node[transform shape=false] at (0,-0.7) {\tiny $\frac{2\pi}{5}$};
    \node[transform shape=false] at (0,1.4) { $A_2$};
    \node[transform shape=false] at (-0.95,-0.9) { $A_1$};
    \node[transform shape=false] at (1,-0.9) { $A_1$};
  \end{tikzpicture}}},\qquad   \qquad
\vcenter{\hbox{\begin{tikzpicture}[scale=1, baseline={(0,-0.1)}]
    \draw[thick, red] (0,0) -- (0,1.2);
    \draw[thick, blue] (0,0) -- ++(198:1.2);
    \draw[thick, blue] (0,0) -- ++(-18:1.2); 
    \draw [thick, <->, >=stealth] (0.0436,0.498) arc[start angle=85,end angle=-13, radius=0.5];
    \draw [thick, <->, >=stealth] (-0.0436,0.498) arc[start angle=95,end angle=193, radius=0.5];
    \draw  [thick, <->, >=stealth]  (0.46,-0.195) arc[start angle=-23,end angle=-157, radius=0.5];
    \node at (0,1.4) { $A_1$};
    \node at (-1.1,-0.7) { $A_2$};
    \node at (1.1,-0.7) { $A_2$};
    \node at (0,-0.7) {\tiny $\frac{4\pi}{5}$};
    \node at (0.7,0.3) {\tiny $\frac{3\pi}{5}$};
    \node at (-0.7,0.3) {\tiny $\frac{3\pi}{5}$};
    \node at (0,-0.2) {\tiny $i$};
  \end{tikzpicture}}}
  \end{equation}
where the angles are fixed by the poles of the $S$-matrix, and the label ``$i$'' indicates that the corresponding scattering amplitude is purely imaginary. Combining the two building blocks in \eqref{fabc}, we obtain an effective 
five-particle scattering of $A_1$:
\begin{equation}
\begin{tikzpicture}[baseline={(0,-0.1)}]
  \draw[thick, red] (0,0) -- (0,1);
  \draw[thick, blue] (0,0) -- ++(198:1); \draw[thick, blue] (0,0) -- ++(-18:1); 
  \draw[thick, red] (-0.961,-0.309) -- ++(-54:1);\draw[thick, red] (-0.961,-0.309) -- ++(234:1);
    \draw[thick, red] (0.961,-0.309) -- ++(-54:1);\draw[thick, red] (0.961,-0.309) -- ++(234:1);
      \node at (0,-0.25) { $i$};
 \end{tikzpicture}
 \quad \Longrightarrow  \qquad 
  \begin{tikzpicture}[baseline={(0,0.05)}]
  \draw[thick, red] (0,0) -- (0,1.17);
  \draw[thick, red] (0,0) -- ++(18:1.17); \draw[thick, red] (0,0) -- ++(162:1.17); \draw[thick, red] (0,0) -- ++(234:1.17); \draw[thick, red] (0,0) -- ++(308:1.17); 
    \node at (0,-0.25) {$i$};
 \end{tikzpicture}
\end{equation}
which is captured by the following identity of the $S$-matrix\footnote{There is an analogous ``cubic" equation in SLYM: $S_{11}(\theta)=S_{11}\left(\theta+\frac{i\pi}{3}\right)S_{11}\left(\theta-\frac{i\pi}{3}\right)$, which admits the minimal solution $S_{11}(\theta)=F_{\frac{2}{3}}(\theta)$ \cite{Cardy:1989fw,Cardy:2023lha}.}
\begin{equation}
    S_{11}(\theta)=S_{11}\left(\theta+\frac{i\pi}{5}\right)S_{11}\left(\theta-\frac{i\pi}{5}\right)S_{11}\left(\theta+\frac{3i\pi}{5}\right)S_{11}\left(\theta-\frac{3i\pi}{5}\right)\,.
\end{equation}
This relation can be derived from the bootstrap equations.

Comparing the amplitudes with the OPE structure of $M(2,7)$, we can identify $A_a\leftrightarrow \phi_{1,1+a}$ \cite{Freund:1989jq}. Using our field identification, $A_a$ can be further identified as $ \phi^a$. The ``$\phi^5$-property" of the fundamental particle $A_1\leftrightarrow\phi$ supports our conjecture (\ref{eq:GL27}).

\section{RG flows between minimal models}
\subsection{Semiclassical argument for the RG flow $M(4,5)\rightarrow M(2,7)$}\label{BCchain}
All of the known RG flows involving $M(2,2n+1)$ are not short, so standard conformal perturbation theory can not be applied. To describe the flow $M(3,4)\rightarrow M(2,5)$ qualitatively, Fisher used a semiclassical argument \cite{Fisher:1978pf}.

Let us show how to generalize Fisher's argument to the flow $M(4,5)\rightarrow M(2,7)$, following a suggestion in \cite{vonGehlen:1994rp}. The tri-critical Ising model $M(4,5)$ perturbed by relevant primary operators is described by
\begin{equation}
    \mathcal{L}=\frac{1}{2}(\partial_\mu\phi)^2+ig_1\phi+g_2\phi^2+ig_3\phi^3+g_4\phi^4+g_6\phi^6\,.
\end{equation}
After shifting the field, $\phi\rightarrow\phi+i\phi_0$, and choosing $\phi_0=\pm \sqrt{\frac{g_4}{15 g_6}}$, the quartic term is removed. Then we find
\begin{equation}
\begin{aligned}
    \mathcal{L}&=\frac{1}{2}(\partial_\mu\phi)^2+i(g_1+2g_2\phi_0-3g_3\phi_0^2-4g_4\phi_0^3+6g_6\phi_0^5)\phi+(g_2-3g_3\phi_0-6g_4\phi_0^2+15g_6\phi_0^4)\phi^2+\\
    &+i(g_3+4g_4\phi_0-20g_6\phi_0^3)\phi^3+6ig_6\phi_0\phi^5+g_6\phi^6+C\,.
\end{aligned}
\end{equation}
Now, by tuning three parameters $g_1=\pm\sqrt{\frac{64g_4^5}{84375g_6^3}}$, $g_2=-\frac{g_4^2}{5g_6}$ and $g_3=\mp\sqrt{\frac{64g_4^3}{135g_6}}$, we obtain the quintic theory (\ref{eq:GL27}), which is conjectured to describe $M(2,7)$, while the operator $\phi^6$ becomes irrelevant in the IR theory.

While this semiclassical argument is expected to be reliable slightly below $d=3$, where both the UV and IR fixed points are quite weakly coupled, it cannot be readily applied in 2D. In this strongly coupled case, one likely needs to resort to numerical methods. The RG flow $M(4,5)\rightarrow M(2,7)$ was studied in \cite{Lencses:2022ira} using the Truncated Conformal Space Approach (TCSA). In \cite{vonGehlen:1994rp} the RG flow $M(4,5)\rightarrow M(2,7)$ was discussed in the context of the quantum Blume-Capel (BC) spin chain \cite{Gefen:1981zz} with a non-Hermitian Hamiltonian:
\begin{equation}\label{eq:BC}
    H_{\text{BC}}=\sum_{i=1}^N (\alpha(S_i^z)^2-S_i^zS_{i+1}^z+h_x S_i^x -ih_zS_i^z)\,,
\end{equation}
where $S_i^x$ and $S_i^z$ are spin-1 matrices. The spectrum of (\ref{eq:BC}) is invariant under $h_x\rightarrow-h_x$. The last term breaks $\mathbb{Z}_2$-symmetry acting as $S_i^z\rightarrow -S_i^z$ but preserves $\mathcal{PT}$-symmetry $S_i^z\rightarrow -S_i^z$, $i\rightarrow -i$. In the space of $\{\alpha,h_{x},h_{z}\}$, $M(4,5)$ is a point, $M(3,4)$ is a line and $M(2,5)$ is a 2D surface.

In a $1+1$ dimensional lattice Hamiltonian approach, the operator $(Q_2+ \bar{Q}_2)\phi_{1,3}$ (\ref{eq:spin2}) is allowed in the IR $M(2,7)$. Therefore, we expect that there are three tunings required to reach the $M(2,7)$ critical point corresponding to the relevant operators $\phi_{1,2}$, $\phi_{1,3}$ and $(Q_2+ \bar{Q}_2)\phi_{1,3}$\,\footnote{In \cite{vonGehlen:1994rp} it was stated that we need only two tunings and the critical $M(2,7)$ manifold $\{\alpha_*,h_{x,*},h_{z,*}\}$ is a line. We instead expect it to be a point.}. However, in the classical Blume-Capel model with an imaginary magnetic field on a 2D square lattice \cite{PhysRevLett.84.4794}, 
\begin{equation}\label{eq:BCclass}
    E_{\text{BC}}=-\sum_{\langle ij\rangle } \sigma_i \sigma_j + \sum_i (\alpha\sigma_i^2-ih_z\sigma_i)\,,
\end{equation}
where $\sigma_i\in\{-1,0,1\}$, the discrete rotation symmetry forbids $T'_{\mu\nu}$ because it changes sign under a rotation by $\pi/2$. Therefore, in this case, only two tunings should be needed to reach $M(2,7)$. This is one more tuning than what was required to reach $M(2,5)$ from $M(3,4)$, so $M(2,7)$ is a candidate to be the tri-critical Yang-Lee model \cite{Lencses:2022ira}. 
\subsection{Flow $M(2,7)\rightarrow M(2,5)$}
Let us study the flow from the quintic theory to the cubic theory $M(2,7)\rightarrow M(2,5)$, clarifying the discussion of \cite{Lencses:2023evr}. It is necessary to perturb the UV CFT by two operators, $i\phi_{1,2}$ and $\phi_{1,3}$. Therefore, the GL description of such a scaling $M(2,7)$ minimal model is
\begin{equation}
    \mathcal{L}=\frac{1}{2}(\partial_\mu\phi)^2+ig_1\phi+g_2\phi^2+ig_5\phi^5\,.
\end{equation}
After the shift $\phi\rightarrow\phi+i\phi_0$ and choosing $\phi_0=-\sqrt[3]{\frac{g_2}{10g_5}}$, the quadratic term is removed. Then we find 
\begin{equation}
    \mathcal{L}=\frac{1}{2}(\partial_\mu\phi)^2+i(g_1-15g_5\phi_0^4)\phi-10ig_5\phi_0^2\phi^3-5g_5\phi_0\phi^4+ig_5\phi^5+C\,.
\end{equation}
By tuning only one parameter $g_1=\sqrt[3]{\frac{27g_2^4}{80g_5}}$, we obtain cubic Yang-Lee theory. The operators $\phi^4$ and $i\phi^5$ become irrelevant in the IR theory.

While this semiclassical argument is suggestive, it is not parametrically reliable. Even in 3D, where the UV CFT is weakly coupled due to the proximity to the upper critical dimension $10/3$, the IR Yang-Lee CFT is not. In 2D, both the $i\phi^5$ and $i\phi^3$ CFTs are strongly coupled, but we can use the TCSA to study the RG flow. This approach was introduced in \cite{Yurov:1989yu}, where it was applied to SLYM up to the level $L=5$ (17 spin-$0$ states).\footnote{Recently, the same approach using the basis of quasi-primary fields was applied to SLYM and perturbed $D_6$ series of $M(3,10)$ up to the level $L=15$ (957 spin-$0$ states) \cite{Katsevich:2024sov}. Quasi-primary fields can be used to identify operators $\phi^k$ in two-dimensional cubic theory \cite{ArguelloCruz:2025zuq}.} Here we are using TCSA with the basis of all descendants excluding null-vectors up to the level $L=25$ (523953 spin-$0$ states). We use the structure constants $C_{223}\approx 2.569$, $C_{233}\approx 4.592i$, $C_{333}\approx -6.019$.

The Hamiltonian of perturbed $M(2,7)$ on an infinite cylinder with circumference $R$ is
\begin{equation}
    H=\frac{2\pi}{R}\left(L_0+\bar{L}_0-\frac{c(2,7)}{12}\right)+i\lambda_{12}\int_0^Rdx\phi_{1,2}(x,0) -\lambda_{13}\int_0^Rdx\phi_{1,3}(x,0)\,.
\end{equation}
This scaling theory has a single scaling parameter $\xi=\frac{\lambda_{12}}{|\lambda_{13}|^{9/10}} {\rm sgn} (\lambda_{13})$, which should be tuned to its critical value to obtain the flow $M(2,7)\rightarrow M(2,5)$.

Acting on descendants of scaling dimension $\Delta=h+\bar{h}$, the TCSA perturbed Hamiltonian  \cite{Yurov:1989yu} takes the following form:
\begin{equation}
    H=\frac{2\pi}{R}\left(\Delta+\frac{17}{21}\right)\bm{1}+i\frac{2\pi G_{12}}{R}B_{12} - \frac{2\pi G_{13}}{R}B_{13}~.
\end{equation}
In this Hamiltonian, $G_{1n}=\lambda_{1n}(2\pi)^{\Delta_{1,n}-1}R^{2-\Delta_{1,n}}=\lambda_{1n,0}(2\pi)^{\Delta_{1,n}-1}r^{2-\Delta_{1,n}}$  are dimensionless effective coupling constants, and $B_{1n}$ are dimensionless matrix elements of $\phi_{1,n}$. $\lambda_{1n,0}$ and $r$ are defined by introducing an arbitrary mass scale $m$, namely $\lambda_{1n,0}=\lambda_{1n} m^{\Delta_{1,n}-2}$ and $r = mR$, where $r$ is understood as the dimensionless  scaling length. Both perturbations $M(2,7)+i\phi_{1,2}$ and $M(2,7)+\phi_{1,3}$ are integrable massive deformations with the spectrums consisting of two breathers. In both cases, we denote the mass of the lighter breather by $m_1$. The couplings $\lambda_{12}$ and $\lambda_{13}$ satisfy the relations 
\cite{Fateev:1993av,Zamolodchikov:1995xk,Fateev:1997yg}:
\begin{equation}
    \lambda_{12}\approx0.07856m_1^{\frac{18}{7}},\quad\lambda_{13}\approx -0.04054m_1^{\frac{20}{7}}\,.
\end{equation}
Therefore, it is convenient to choose $m=m_1$ as the reference mass scale in the TCSA Hamiltonian. The same choice was also used in \cite{Lencses:2023evr}.

\begin{figure}[ht]
    \centering
    \begin{subfigure}{0.48\linewidth}
        \centering
        \includegraphics[width=\linewidth]{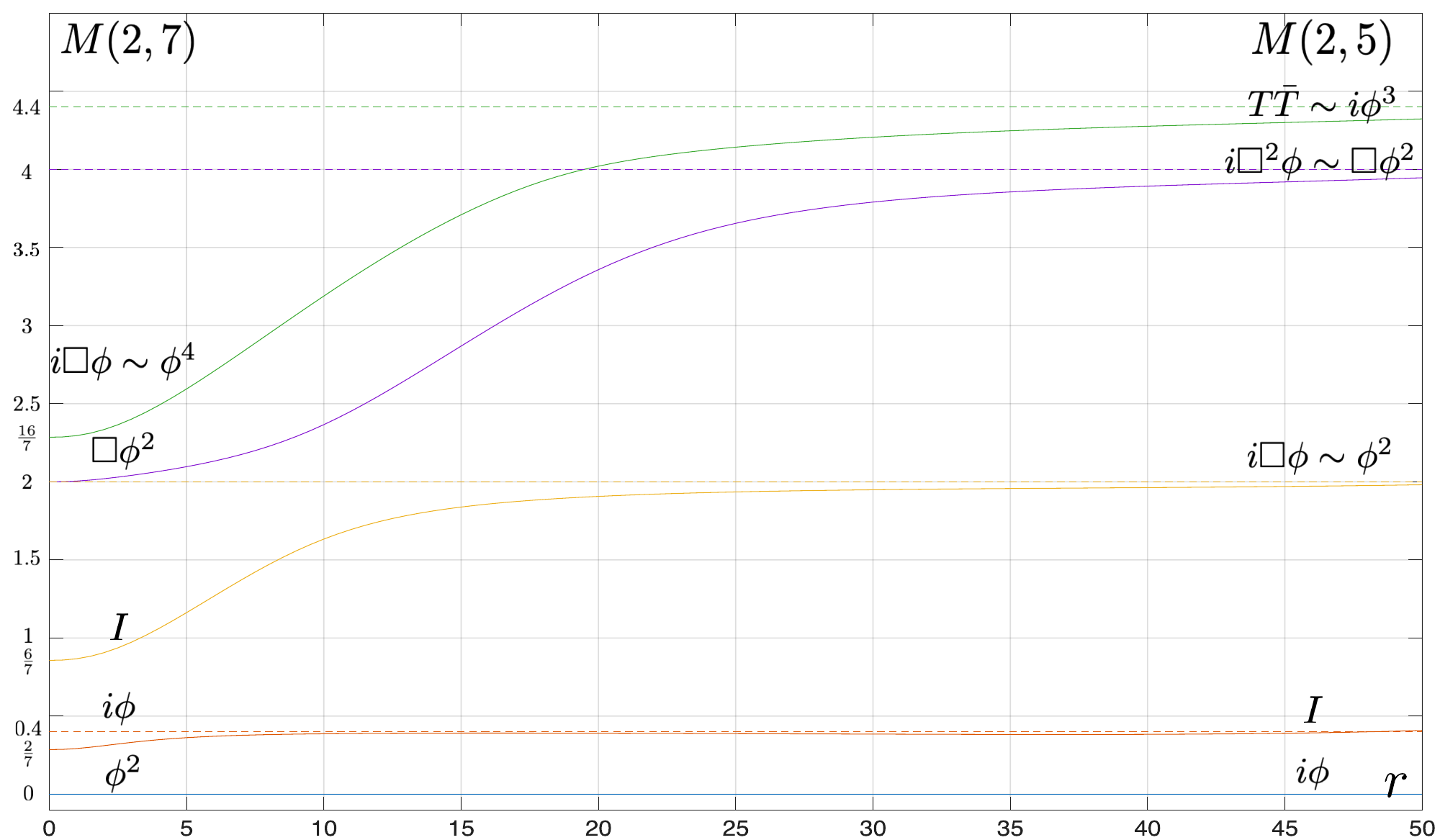}
        \caption{$-\frac{C_n(r)-C_0(r)}{12}$.}
        \label{fig:Cn-C0}
    \end{subfigure}%
    \hfill
    \begin{subfigure}{0.48\linewidth}
        \centering
        \includegraphics[width=\linewidth]{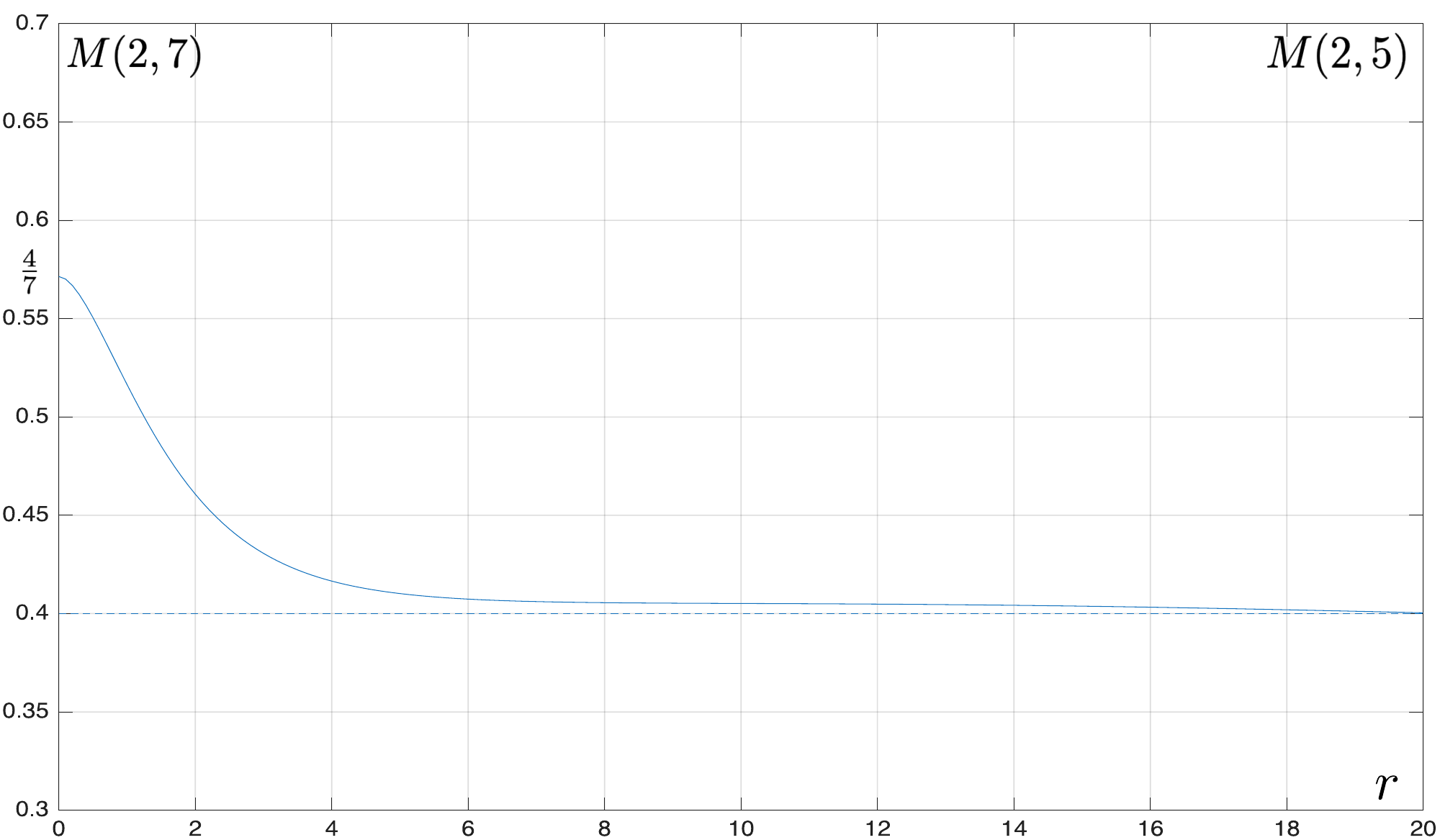}
        \caption{$C_0(r)$ between $c_{\text{eff}}^{\text{UV}}=\frac{4}{7}$ and $c_{\text{eff}}^{\text{IR}}=\frac{2}{5}$.}
        \label{fig:C0}
    \end{subfigure}
    \caption{Operator flow (a) and effective central charge (b) 
    for the $M(2,7)\rightarrow M(2,5)$ flow.}
    \label{fig:combined}
\end{figure}

The energy of the $n$-th energy level on a cylinder is \cite{Bloete:1986qm, Affleck:1986bv,Itzykson:1986pk}:
\begin{equation}
    E_n(R)=FR-\frac{C_n(R)}{12}\frac{2\pi}{R}\,,
\end{equation}
where $F$ is the bulk vacuum energy density and $C_n(R)$ interpolates between  $c^{\text{UV}}-12\Delta^{\text{UV}}_n$ at $R\rightarrow0$ and $c^{\text{IR}}-12\Delta^{\text{IR}}_n$ at $R\rightarrow\infty$. Applying TCSA, we plot the first 5 energy levels $-\frac{C_n(R)-C_0(R)}{12}=\frac{R}{2\pi}(E_n(R)-E_0(R))$ interpolating between $\Delta_n^{\text{UV}}-\Delta_0^{\text{UV}}$ and $\Delta_n^{\text{IR}}-\Delta_0^{\text{IR}}$ for tuned $\xi$ (Fig. \ref{fig:Cn-C0}). We also plot $C_0(R)=-\frac{6R}{\pi}(E_0(R)-FR)$, which  interpolates between $c_{\text{eff}}^{\text{UV}}=c_{\text{eff}}(2,7)=\frac{4}{7}$ and $c_{\text{eff}}^{\text{IR}}=c_{\text{eff}}(2,5)=\frac{2}{5}$ (Fig. \ref{fig:C0}) as a function of the  dimensionless parameter $r=m_1R$,  for appropriately  tuned $\xi$ and $F$. For the flow $M(2,7)\rightarrow M(2,5)$, we obtained:
\begin{equation}
    \xi=1.28855,\quad F=-0.093535\,.
\end{equation}
Our $\xi$ differs from \cite{Lencses:2023evr} by $0.05\%$. For the flow with the same IR Yang-Lee model $M(3,4)\rightarrow M(2,5)$, in \cite{Xu:2022mmw} it was obtained free energy $F=0.092746$, which magnitude differs only by $0.8\%$. 

From the TCSA plot (Fig. \ref{fig:Cn-C0}) and our field identification (Table. \ref{tab:2}), we can obtain operator flow from $M(2,7)$ to $M(2,5)$\footnote{The operator flow from $M(3,4)$ to $M(2,5)$ was obtained in \cite{ArguelloCruz:2025zuq}. In particular, the RG flow interchanges the identity operator and $\phi$, i.e. $I\to i\phi, i\phi\to I$. By contrast, in 3D, the flow from Ising CFT to Yang-Lee CFT preserves the identity operator, i.e. $\phi^k\to \phi^k$ for $k=0,1,2,3$. \cite{ArguelloCruz:2025zuq} }. Spin $0$: $\phi^2\rightarrow i\phi$, $i\phi\rightarrow I$, $I\rightarrow i\Box\phi\sim\phi^2$, $\Box\phi^2\rightarrow i\Box^2\phi$, $i\Box\phi\sim \phi^4\rightarrow T\bar{T}\sim i\phi^3$. Spin 1: $\partial\phi^2\rightarrow i\partial\phi$, $i\partial\phi\rightarrow i\partial\Box\phi$. We expect in 3D the operator flow from $i\phi^5$ to $i\phi^3$ to be: $\phi^k\rightarrow \phi^k$ in accordance with Fig. \ref{Padeplot}. 

\section{Discussion}
We have presented evidence that the GL description of $M(2,7)$ is provided by the action (\ref{eq:GL27}). The upper critical dimension of this $i\phi^5$ universality class is $10/3$, so the 3D theory is rather weakly coupled and this is supported by Fig. \ref{Padeplot}. While the scaling dimensions of operators appear to be close to the free dimensions, finding their precise values is an interesting problem.
It would also be very interesting to calculate the scaling dimensions numerically in a lattice model, perhaps following the non-Hermitian Hamiltonian approach of \cite{vonGehlen:1994rp}, which used the Blume-Capel spin chain in an imaginary magnetic field.
We hope that our estimates in $d=3$ can be compared with numerical quantum criticality in the corresponding model on regularized spheres.

Let us note that $M(2,2n+1)$ minimal models have $\mathcal{PT}$ symmetry implied by OPE structure and no $\mathbb{Z}_2$ symmetry \cite{Lassig:1991an}. This also holds for scalar theories with odd potentials $i\phi^{2n-1}$, where $\phi\rightarrow-\phi$ is not a symmetry but $\phi \rightarrow -\phi$, $i\rightarrow -i$ is a symmetry. 
Therefore, we conjecture that the general GL description of $M(2,2n+1)$ minimal models is
\begin{equation}\label{eq:GLgen}
    S_{2,2n+1}=\int d^dx\left(\frac{1}{2}(\partial_\mu\phi)^2+\frac{g\phi^{2n-1}}{(2n-1)!}\right),\quad g\in i\mathbb{R}\,.
\end{equation}
The OPEs appear to work, but we leave the details of this conjecture for further work.

\section*{Acknowledgements}
We are grateful to A. M. Polyakov and A. B. Zamolodchikov for very useful discussions. 
This work was supported in part by the US National Science Foundation Grant No.~PHY-2209997 and by the Simons Foundation Grant No.~917464. Z.S. is supported in part by the U.S. Department of Energy grant DE-SC0009988. G.T. work was supported in part by the U.S. Department of Energy Grant No. DE-SC0010118 and by the Simons Foundation Grant No.~994316.

\bibliography{refs}
\bibliographystyle{utphys}

\end{document}